\documentclass[12pt,preprint]{aastex}
\begin{document}
\title{Discovery of the Neutron Star Spin Frequency in EXO 0748--676}
\author{Adam R. Villarreal\altaffilmark{1}} \affil{Department of
Physics, University of Arizona, Tucson, AZ 85721;
adamv@physics.arizona.edu} \and \altaffiltext{1}{NASA Harriett G.
Jenkins Fellow} \author{Tod E. Strohmayer} \affil{Laboratory for
High Energy Astrophysics, NASA Goddard Space Flight Center,
Greenbelt, MD 20771; stroh@milkyway.gsfc.nasa.gov}
\begin{abstract}

We report the results of a search for burst oscillations during
thermonuclear X-ray bursts from the low mass X-ray binary (LMXB)
EXO 0748--676.  With the proportional counter array (PCA) onboard
the {\it Rossi X-ray Timing Explorer} (RXTE) we detected a 45 Hz
oscillation in the average power spectrum of 38 thermonuclear
X-ray bursts from this source. We computed power spectra with 1 Hz
frequency resolution for both the rising and decaying portions of
38 X-ray bursts from the public RXTE archive. We averaged the 1 Hz
power spectra and detected a significant signal at 45 Hz in the
decaying phases of the bursts. The signal is detected at a
significance level of $4 \times 10^{-8}$. No similar signal was
detected in the rising intervals.  A fit to the oscillation peak
at 0.25 Hz resolution gives a frequency of $\nu_0 = 44.7 \pm 0.06$
Hz and an oscillation quality factor of $Q = \nu_0 /
\Delta\nu_{fwhm} = 80 \pm 18$. The average signal amplitude is
$\approx 3\%$ (rms). The detection of 45 Hz burst oscillations
from EXO 0748--676 provides compelling evidence that this is the
neutron star spin frequency in this system. We use the inferred
spin frequency to model the widths of absorption lines from the
neutron star surface and show that the widths of the absorption
lines from EXO 0748--676 recently reported by Cottam et al. are
consistent with a 45 Hz spin frequency as long as the neutron star
radius is in the range from about 9.5 -- 15 km.  With a known spin
frequency, precise modelling of the line profiles from EXO
0748--676 holds great promise for constraining the dense matter
equation of state.

\end{abstract}

\section{Introduction}

High time resolution observations of thermonuclear X-ray bursts
with the \textit{Rossi X-ray Timing Explorer} (\textit{RXTE}) have
now found X-ray brightness oscillations, or burst oscillations,
from more than a dozen LMXBs. A large body of evidence supports
the conclusion that such oscillations are produced by spin
modulation of the X-ray burst flux (for a recent review see
Strohmayer \& Bildsten 2003 and references therein).  In the last
two years the detection of burst oscillations at the known spin
frequencies of two accreting millisecond pulsars; SAX
J1808.4--3658 (Chakrabarty et al. 2003), and XTE J1814--338
(Strohmayer et al. 2003) has provided particularly strong
confirmation of the spin modulation hypothesis.

EXO 0748--676 (hereafter EXO 0748) is a well studied low-mass
X-ray binary (LMXB) discovered by \textit{EXOSAT} in 1985 (Parmar
et al.  1985). This transient X-ray burst source exhibits
irregular X-ray dips and periodic eclipses (Gottwald et al. 1986).
Timing of the eclipses revealed a 3.82 hr orbital period (Parmar
et al. 1986; Hertz et al. 1996; Wolff et al. 2002). Based on the
eclipse duration and orbital period the system inclination angle
is constrained to be in the range 75--82$\degr$ (Parmar et al.
1986), which depends on assumptions about the neutron star's
companion. X-ray timing studies of the source have revealed the
presence of $\approx$1 Hz quasi-periodic oscillations (QPOs), as
well as a single kilohertz QPO, which is presumably the lower kHz
peak of a pair (Homan et al. 1999; Homan \& van der Klis 2000).
Homan \& van der Klis (2000) searched a total of 10 X-ray bursts
from EXO 0748 for burst oscillations. They searched the 100--1000
Hz frequency band (with 2 Hz resolution), but found no significant
signals. They placed upper limits on the amplitude of oscillations
to be between 4 \% and 11 \% (rms) during the rise of the bursts
in the 2--60 keV band.

EXO 0748 is one of few neutron star LMXBs for which
high-resolution spectroscopic observations have been obtained for
a large number of X-ray bursts. Cottam et al. (2002) reported the
presence of narrow absorption lines in a study of EXO 0748 burst
spectra observed with the Reflection Grating Spectrometer (RGS) on
\textit{XMM--Newton}. Using co-added data from 28 bursts, they
found evidence for absorption features between 13--14 $\AA$ which
they interpreted as redshifted absorption lines of the n = 2 to 3
transitions from hydrogen- and helium-like Fe.  Their line
identifications imply a neutron star surface redshift of $z =
0.35$. Here, $1 + z = (1 - 2GM/c^2 R)^{-1/2}$, where $M$ and $R$
are the neutron star mass and radius, respectively. A measurement
of $z$ fixes the mass to radius ratio, or the compactness, but
additional information is required to determine both the mass and
radius separately.

A measurement of the neutron star spin frequency in EXO 0748 could
provide the additional information required to measure both its
mass and radius. This is possible because the observed width of
surface spectral lines provides information about the surface
rotational velocity, $v_{rot}$, and thus the stellar radius if the
spin frequency, $\nu_{spin}$, is known, viz. $v_{rot} \propto
\nu_{spin} R\,$. Even at modest rotation rates it is expected that
rotational Doppler broadening will dominate over thermal Doppler
and Stark broadening (see Bildsten, Chang \& Paerels 2003).
Rotational broadening depends on the surface rotational velocity,
$v_{rot} = 2\pi \nu_{spin} R \sin i$, where $i$ is the system
inclination (assuming the rotation axis is perpendicular to the
orbital plane). Since the inclination of EXO 0748 is tightly
constrained by the presence of eclipses, a measurement of the
stellar radius would, in principle, be possible if the spin
frequency were known. A radius determination, combined with $z$,
would then allow for a constraint to be placed on the mass. Until
now, this has not been possible because burst oscillations and
surface absorption lines have not been observed in the same
source. Precise neutron star mass and radius measurements are
crucial for understanding the equation of state (EOS) of
supranuclear density matter (see, for example, Lattimer \& Prakash
2001; Strohmayer 2004).

In this Letter we report the discovery of 45 Hz burst oscillations
from EXO 0748 with the proportional counter array (PCA) onboard
\textit{RXTE}. We investigate the implications of this finding for
the widths of absorption lines from the neutron star surface. In
\S 2 we summarize our search and detection of a 45 Hz burst
oscillation signal in the average power spectrum of the decay
phases of 38 bursts from EXO 0748. In \S 3 we show that the width
of the lines observed from EXO 0748 with the RGS are consistent
with a 45 Hz spin frequency if the neutron star radius is 9.5--15
km. We conclude in \S 4 with a brief summary of our results.

\section{Observations and Data Analysis}

Since oscillation searches in individual bursts from EXO 0748 have
been unsuccessful, we elected to search by ``stacking''
(averaging) the power spectra from all available bursts. We began
by searching the public \textit{RXTE} data archive and found a
total of 38 type I X-ray bursts from EXO 0748. All the bursts were
observed (between 15 August, 1996 and 19 February, 2003) with the
PCA and high time resolution event mode data were available. For
the purposes of computing power spectra we used light curves
sampled at 4096 Hz, yielding a Nyquist frequency of 2048 Hz. Since
burst oscillation amplitudes generally increase with photon energy
(see Strohmayer et al. 1997; Muno, \"{O}zel \& Chakrabarty 2003),
we computed power spectra in the energy band 6--60 keV. All power
spectra were normalized such that a pure Poisson noise process
would be flat with a mean of 2 (see Leahy et al. 1983).

From the sample of 38 bursts we computed two average power
spectra; one each for the rises and decays of the burst profiles.
We selected rising intervals from just prior to burst onset to
near the burst peak. We started the decay intervals at the end of
the rise intervals and stopped when the countrate had fallen to
about 5\% of the peak rate (above the pre-burst level).  For both
the rises and decays we ``rounded-off'' the intervals so that the
length of each was an even multiple of the shortest length.
Because we have bursts of differing length this allowed us to
linearly rebin the power spectra of each burst to the same
frequency resolution before averaging.  This procedure resulted in
rise intervals lasting 5 or 10 s and decay intervals from 16 to
256 s, with most being 64 or 128 s.

Since burst oscillations can drift in frequency by order 1--2 Hz
we rebinned each individual burst power spectrum to 1 Hz
resolution and then averaged the 1 Hz power spectra of all the
bursts (rises and decays separately).  We estimated the errors on
the individual burst power spectra using the statistical error
associated with rebinning (averaging) $N_i$ independent powers for
the $i^{\rm th}$ burst, viz. $\sigma_i = 2/\sqrt(N_i)$. We then
propagated these uncertainties to estimate the errors for the
averaged spectra. Figure 1 shows the average power spectrum
computed from the decaying portion of the bursts in the 1--2048 Hz
band. This spectrum contains a prominent peak at $\approx$45 Hz.
The increase in mean power below about 20 Hz is not unexpected,
since, by definition, the decay light curves are all trending down
in countrate.

In order to quantify the significance of this peak we need to
understand the noise power distribution of our power spectrum.
First, we fitted the continuum power level from 20--2048 Hz with a
constant + power-law model, excluding the 45 Hz peak to avoid
biasing the fit to higher values. We then rescaled our power
spectrum by dividing by the best fitting continuum model. Figure 2
shows the power spectrum and best fitting continuum model (top)
along with the rescaled spectrum (bottom). To estimate the noise
power distribution we computed a histogram of the number of noise
powers with power between $p_i$ and $p_i+\Delta p$, using $\Delta
p = 0.01$. Figure 3 shows the resulting noise power histogram and
best fitting $\chi^2$ distribution (solid). A $\chi^2$
distribution with a mean of 2 and 3,446 degrees of freedom fits
extremely well. This distribution is, to high accuracy, a Gaussian
with a mean of $2.000 \pm 0.0011$ and a standard deviation of
$\sigma=0.0481 \pm 0.0008$. It is not unexpected that the
distribution is effectively Gaussian, especially given the large
number of independent powers averaged (van der Klis 1989).

We now estimate the significance of the 45 Hz peak using the
fitted $\chi^2$ distribution. The 45 Hz peak has a power value of
2.335 (vertical dashed line in Figure 3), and the single trial
probability of obtaining this value from the fitted distribution
is $2.1 \times 10^{-11}$. Using a conservative number of trials of
2048 (the number of bins in our power spectrum), we arrive at a
significance of $4.3 \times 10^{-8}$, indicating a strong
detection. At 0.25 Hz resolution we could just resolve the peak. A
fit with a Lorentzian function gives a frequency $\nu_0 = 44.7 \pm
0.06$ Hz, yielding a quality factor of $Q = \nu _0 /
\Delta\nu_{fwhm} = 80 \pm 18$.  This peak corresponds to an
average signal amplitude of $\approx 3\%$ (rms).

To further investigate the association of the 45 Hz signal with
the X-ray bursts from EXO 0748 we computed two additional power
spectra in the same manner as before; one for the first half (ie.
the brighter half) of the burst decay intervals and the other for
the second half. These two power spectra are compared in Figure 4.
The power spectrum from the first half (top) shows a strong peak
at 45 Hz while no signal is detected in the other (fainter) half.
This demonstrates that the 45 Hz signal is associated with the
brighter portions of the burst decay profiles, as is typical for
burst oscillations in other sources.  As an additional test of
robustness, we divided the burst sample into two sets (each with
19 bursts) and computed average power spectra for each set.  We
detected the 45 Hz peak in both sets.

\section{The Spin Frequency and Line Profiles}

We have detected an oscillation signal at 45 Hz during X-ray
bursts from EXO 0748.  The frequency width of the signal and its
clear association with the X-ray bursts (see Figure 4) indicates
that it is most likely a burst oscillation signal similar to those
seen in more than a dozen other LMXBs, therefore establishing a
spin frequency of 45 Hz for the neutron star in EXO 0748.
Interestingly, this is the slowest spin period yet measured for a
burst oscillation source. This begs the question, why is the spin
frequency so much lower in EXO 0748 compared to other LMXBs?  One
possibility is that the source is in spin equilibrium with a
neutron star magnetic field that is stronger than in other LMXBs.
Using typical estimates of the accretion torque (see, for example,
Psaltis \& Chakrabarty 1999) one would require a magnetic field
ranging from $B \approx 1.5-5 \times 10^9$ G, assuming a
time-averaged mass accretion rate in the range 0.01--0.1 $\dot
m_{Edd}$, where $\dot m_{Edd}$ is the Eddington accretion rate.
Knowledge of the spin frequency also has implications for the kHz
QPO frequencies expected from the system. Based on behavior from
other sources (ie. those with $\nu_{spin} < 400$ Hz), one would
expect a frequency separation of 45 Hz for the kHz QPO pair. This
would make a nice test case; however, the comparatively small
frequency separation (comparable to the width of the lower kHz QPO
in some systems) could make resolving both peaks difficult.

The detection of the spin frequency in EXO 0748 is very important
given the evidence for narrow, gravitationally redshifted
absorption lines from this object (Cottam et al. 2002). To
investigate the consistency of the absorption line widths observed
by Cottam et al. (2002) with a neutron star spin frequency of 45
Hz, we computed model line profiles from a rotating neutron star
(see also, Datta \& Kapoor 1988; {\" O}zel \& Psaltis 2003). Our
modelling builds on that previously described by Nath, Strohmayer
\& Swank (2002) and is discussed in detail by Strohmayer (2004).
Briefly, the model includes photon deflection in the Schwarzschild
metric, relativistic beaming and aberration, gravitational
redshifts, and allows for arbitrary viewing geometries. We assume
the whole surface of the neutron star is involved in line
formation, and that the rotational axis is perpendicular to the
orbit plane. We use an inclination of $78.5 \degr$. We model the
intrinsic line shape as a Gaussian with a width (FWHM) fixed at
the value due to Stark and thermal Doppler broadening estimated by
Bildsten, Chang \& Paerels (2003) for the H$\alpha$ transition.
Finally, we convolve the model line profile observed at infinity
with an RGS response model appropriate for the 13--14 $\AA$ band.

A detailed model fit to the RGS absorption line data is beyond the
scope of this paper; however, we can compare the measured line
width with our line broadening model and place some preliminary
limits on the stellar radius. To do this we used data from Cottam
et al. (2002, Figure 1) to model the widths of the absorption
lines at 13 and 13.75 $\AA$. We find, from a joint fit using
Gaussian profiles and only one width parameter for both lines, a
value of $\Delta\lambda / \lambda_{fwhm} = 0.018 \pm 0.004\,$.
Figure 5 shows absorption line profiles for a neutron star
spinning at 45 Hz with different radii. As expected, larger
neutron stars produce broader absorption lines, and even at 45 Hz
the spin Doppler broadening dominates. At the half-intensity value
of each line profile in Figure 5 is a horizontal line. The thin
part of each line represents the maximum absorption line width,
while the thick part denotes the minimum line width (both at
$1\sigma$). The vertical dotted lines denote the best-fit value of
$\Delta\lambda / \lambda_{fwhm} = 0.018\,$. One can see from
Figure 5 that radii in the range $9.5 < R < 15$ km (or, assuming
$z$ is known, masses in the range $1.5\, M_{\sun} < M < 2.3\,
M_{\sun}$) are consistent with the derived line widths, and that
the observed lines are best matched with $R \approx 11.5$ km ($M
\approx 1.8\, M_{\sun}$).

This comparison demonstrates that the observed line widths are
consistent with a 45 Hz neutron star spin frequency and a
reasonable range of neutron star radii. It also shows that precise
measurements of the absorption line widths can, in principle, lead
to an accurate measurement of the stellar radius.  We will attempt
to derive more precise radius constraints from detailed model
fitting in future work.

\section{Summary}

Our discovery of burst oscillations from EXO 0748--676 has
established a spin frequency of 45 Hz for the accreting neutron
star in this system. We have also shown that the observed widths
of the absorption lines claimed by Cottam et al. (2002) to be
gravitationally redshifted lines from the neutron star surface are
consistent with such a spin frequency and with reasonable neutron
star radii. This provides at least indirect support for the idea
that the absorption lines {\it could} indeed come from the neutron
star surface (ie. they are {\it not} too narrow).  Finally, we
have shown that detailed modelling of absorption line profiles
combined with spin measurements will likely provide a means to
accurately measure both the masses and radii of neutron stars, and
thus tightly constrain the dense matter EOS.

A. R. V. is grateful for the NASA Harriett G. Jenkins Pre-doctoral
Fellowship Program Mini-Grant Award in support of this work.

\pagebreak

\begin{figure}
\begin{center}
\includegraphics[width=6in, height=6in]{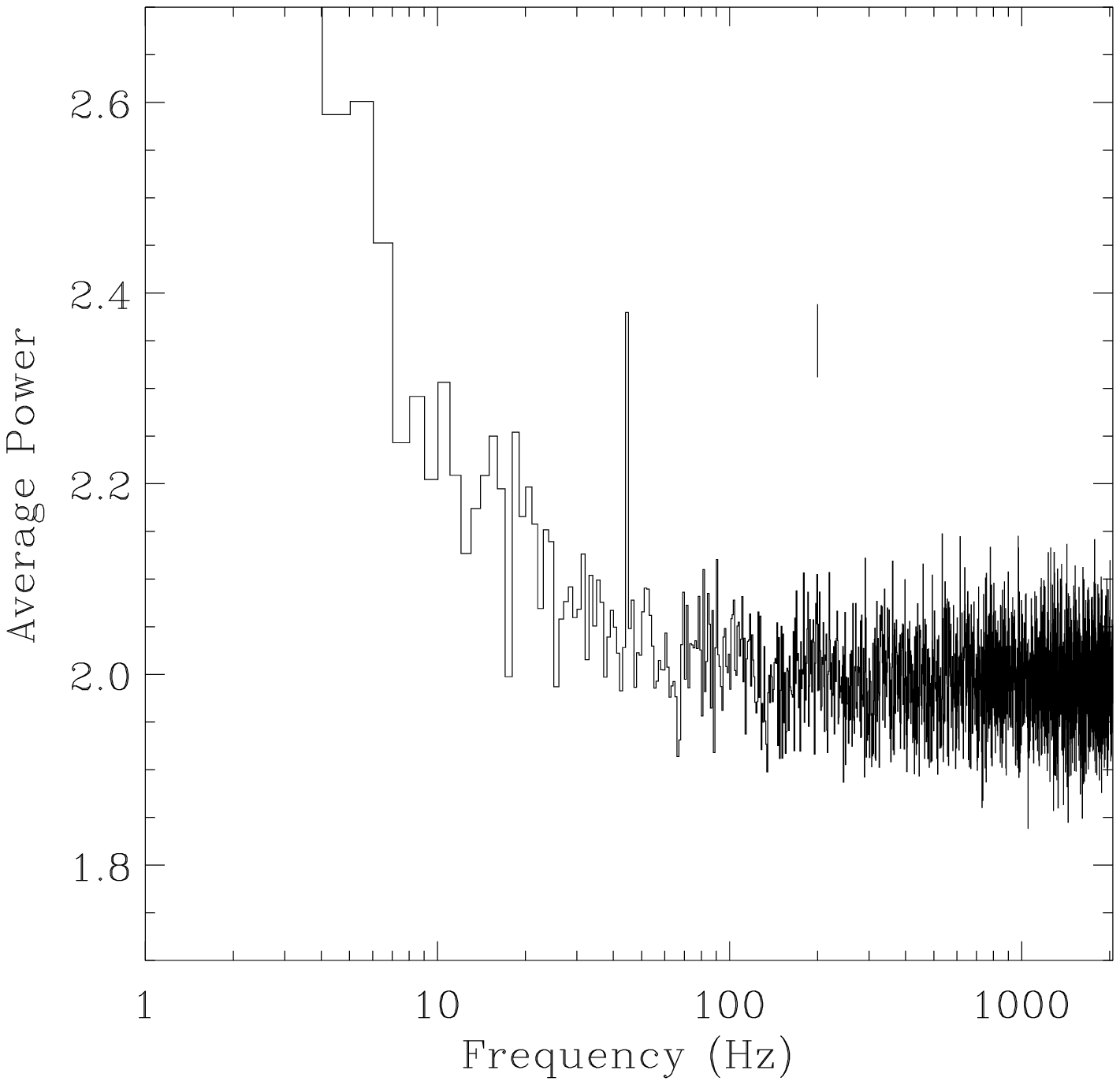}
\end{center}
Figure 1: Average Leahy-normalized power spectrum of the decay
intervals of 38 X-ray bursts from EXO 0748 in the 1--2048 Hz band.
The frequency bins are 1 Hz and the Nyquist frequency is 2048 Hz.
Note the prominent peak at $\approx$45 Hz.  The increase in power
towards low frequencies is due to the decrease in countrate with
time during the burst decays. A characteristic error bar is also
shown.

\end{figure}
\clearpage

\begin{figure}
\begin{center}
\includegraphics[width=6in,
height=6in]{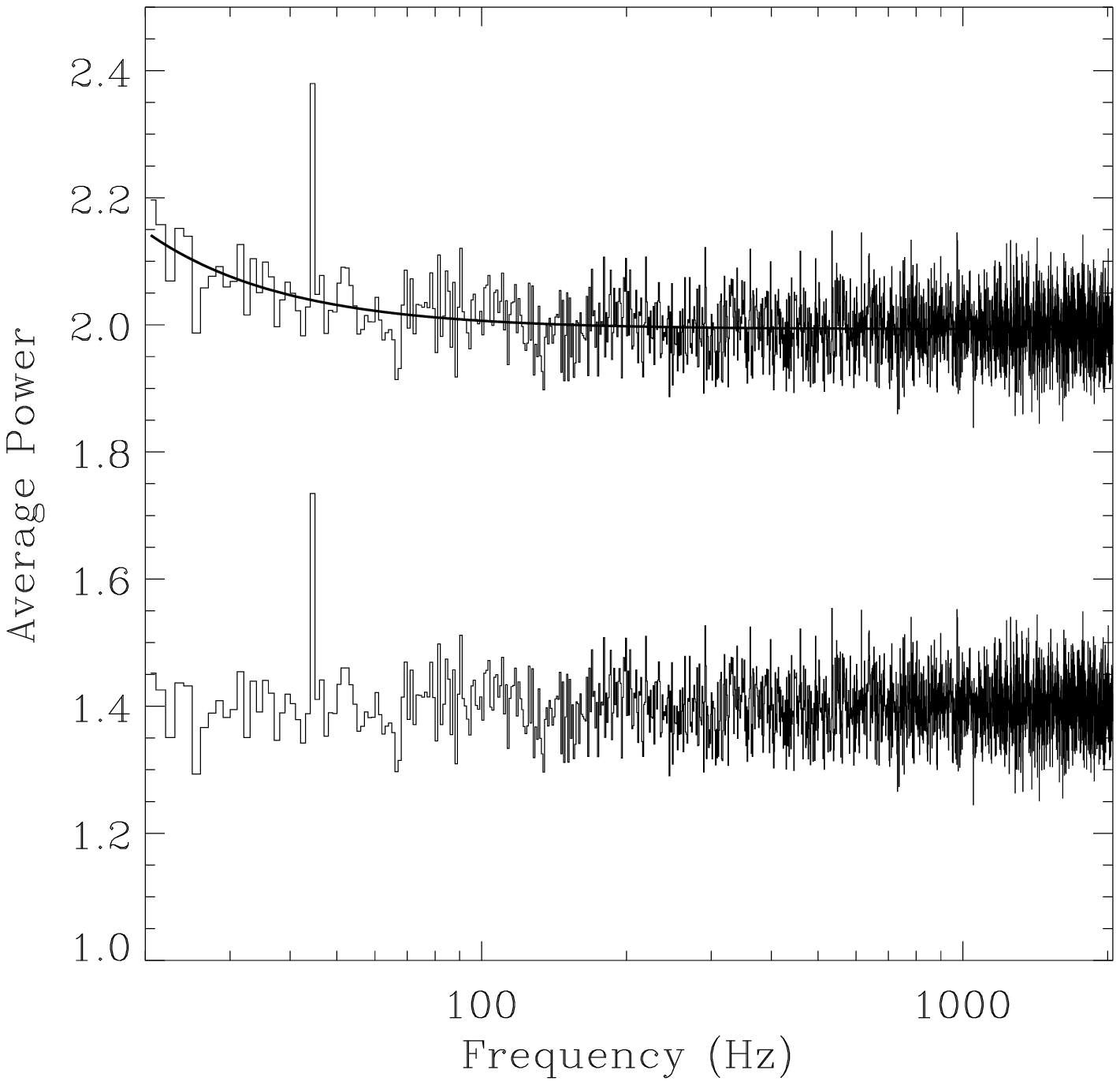}
\end{center}
Figure 2: The same power spectrum from Figure 1 is shown in the
20--2048 Hz band along with the best fitting constant + power law
model of the continuum (top).  The renormalized power spectrum
obtained by dividing by the continuum model (and multiplying by 2)
is also shown (bottom). The renormalized spectrum has been
displaced by $-0.6$ for clarity.
\end{figure}
\clearpage

\begin{figure}
\begin{center}
\includegraphics[width=6in, height=6in]{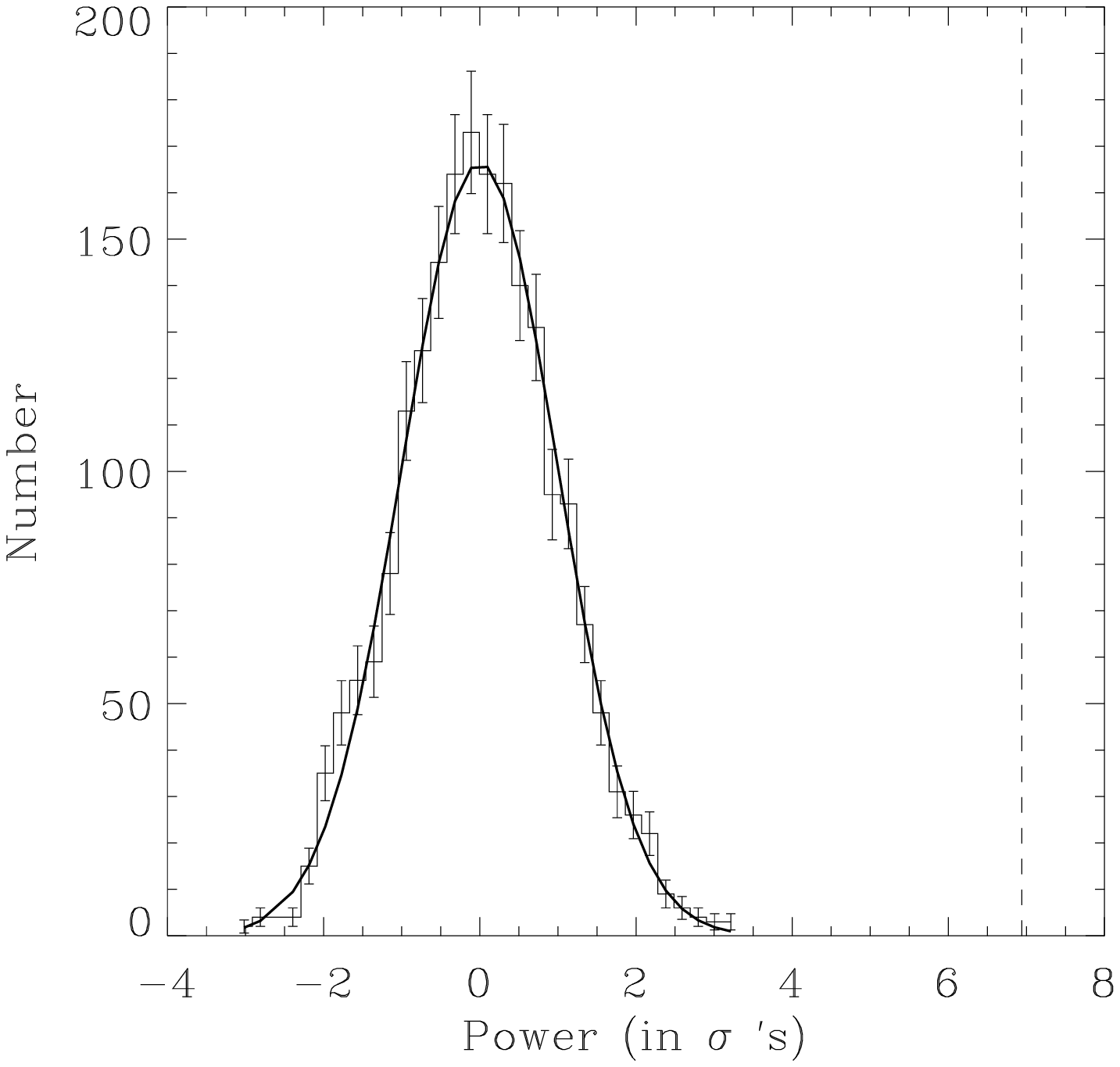}
\end{center}
Figure 3: Distribution of noise powers for the average
renormalized power spectrum shown in Figure 2 (histogram). The
best fitting $\chi^2$ model is also shown (solid). The
distribution has been scaled to zero mean and the ordinate is in
units of standard deviations ($\sigma$'s).  The power level of the
45 Hz peak is marked by the vertical dashed line and is
$\approx$7$\sigma$ from the mean.
\end{figure}
\clearpage

\begin{figure}
\begin{center}
\includegraphics[width=6in, height=6in]{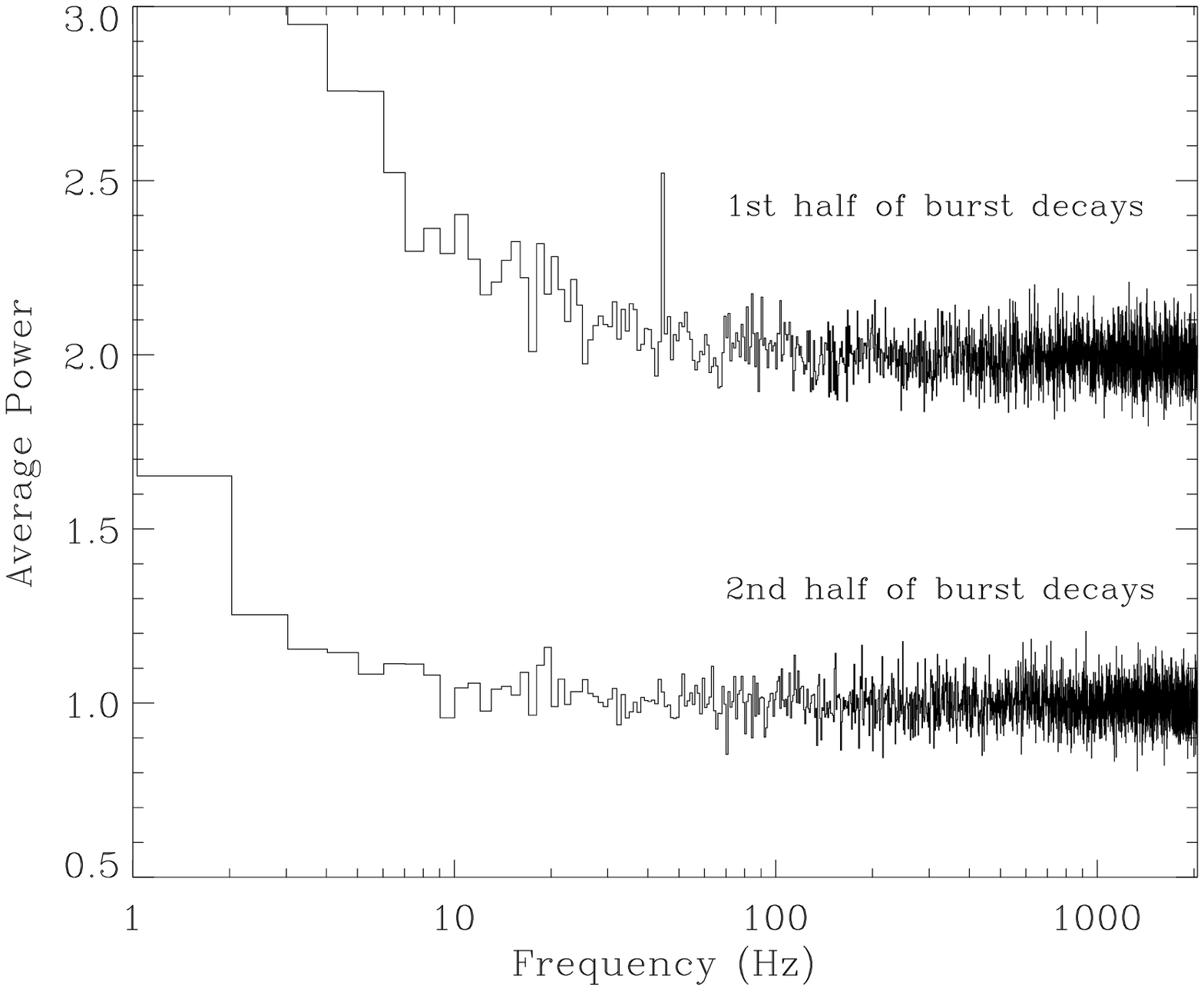}
\end{center}
Figure 4: Comparison of average power spectra computed from the
1st half (by time) of the burst decay intervals (top) and the 2nd
half (bottom). The 45 Hz signal is clearly associated with the 1st
half of the decay intervals.
\end{figure}
\clearpage

\begin{figure}
\begin{center}
\includegraphics[width=6in, height=6in]{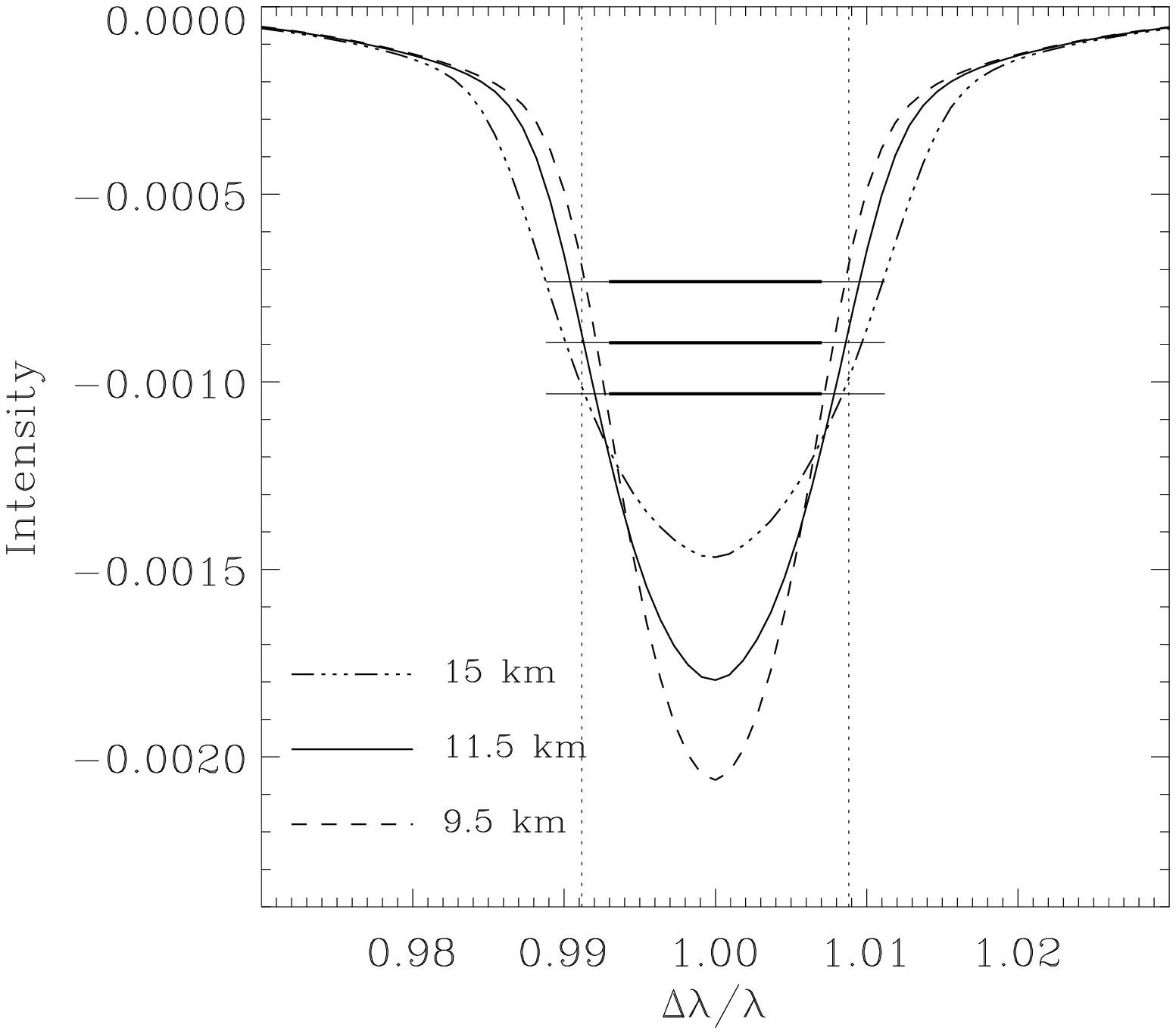}
\end{center}
Figure 5: Model absorption line profiles from a neutron star
rotating at 45 Hz for radii of 9.5 (dashed), 11.5 (solid) and 15
(dot-dashed) km. A horizontal line is located at the
half-intensity point for each profile. The thin portion of each
line represents the maximum width consistent with the RGS data
from Cottam et al. ($1\sigma$), while the thick portion represents
the minimum width (again, at $1\sigma$). The vertical dotted lines
denote the best-fit line width. Comparison of the model profiles
and the measured line widths suggests $R = 11.5^{+3.5}_{-2.5}$.
The model lines have been convolved with a line response function
appropriate for the RGS at 13.5 $\AA$. This accounts for the
shallow but broad wings of the profiles.
\end{figure}
\clearpage

\end{document}